\documentclass[manuscript]{aastex}
\usepackage{emulateapj5}

\newcommand{\lsim}{\raisebox{-0.3ex}{\mbox{$\stackrel{<}{_\sim} \,$}}}

\usepackage{epsfig}
\usepackage{amsmath}
\usepackage{amssymb}
\usepackage{natbib}
\usepackage{graphicx}

\shorttitle{Highly coherent kHz QPO}
\shortauthors{Mukherjee and Bhattacharyya}
\begin{document}

\title{Highly coherent kilohertz quasi-periodic oscillations from a
neutron star X-ray binary EXO 1745--248} % \\

\author{Arunava Mukherjee\altaffilmark{1} and Sudip Bhattacharyya\altaffilmark{1}}
\altaffiltext{1}{Tata Institute of Fundamental Research,
    Mumbai-400005, India; arunava@tifr.res.in; sudip@tifr.res.in}

%\author{C. D. Biemesderfer\altaffilmark{4,5}}
%\affil{National Optical Astronomy Observatories, Tucson, AZ 85719}
%\email{arunava@tifr.res.in}

\begin{abstract}
We report the discovery ($20\sigma$) of kilohertz quasi-periodic oscillations (kHz QPOs)
at $\sim 690$ Hz from the transient neutron star low-mass X-ray binary EXO 1745--248. 
We find that this is a lower kHz QPO, and systematically study the time variation of its
properties using smaller data segments with and without the shift-and-add technique. 
The quality (Q) factor occasionally significantly varies within short ranges of frequency and time.
A high Q-factor ($264.5\pm38.5$) of the QPO is found for a 200 s time segment, 
which might be the largest value reported in the literature. 
We argue that an effective way to rule out kHz QPO models is to observationally
find such high Q-factors, even for a short duration, 
as many models cannot explain a high coherence.
However, as we demonstrate, the shift-and-add technique cannot find a very high Q-factor
which appears for a short period of time. This shows that the coherences of kHz QPOs 
can be higher than the already high values reported using this technique, implying 
further constraints on models. We also discuss the energy dependence of fractional 
rms amplitude and Q-factor of the kHz QPO.
\end{abstract}


\section{Introduction}\label{Introduction}

Many neutron star low mass X-ray binary (LMXB) systems show high frequency
($\sim 200-1200$ Hz) and somewhat coherent intensity variations \citep{vanderKlis2006,
Bhattacharyya2010}. Such variations are known as kilohertz quasi-periodic oscillations 
(kHz QPOs). Sometimes these QPOs appear in a pair; the higher frequency one is called the
upper kHz QPO, while the lower frequency one is known as the lower kHz QPO.
Soon after their discovery \citep{vanderKlisetal1996, Strohmayeretal1996}, it was
realized that they originate from within a few Schwarzschild radii of the neutron star,
and hence could be useful to probe the strong gravity regime, as well as the supranuclear
degenerate matter of the stellar core. Despite this potential, so far kHz QPOs could not
be used as a reliable tool, because their correct theoretical model has not been
identified yet. Many proposed models of this timing feature primarily attempt to 
explain the frequency (e.g., \citet{Milleretal1998, StellaVietri1998, StellaVietri1999,
LambMiller2003, KluzniakAbramowicz2001, AbramowiczKluzniak2001, Wijnandsetal2003, Leeetal2004,
Zhang2004, Mukhopadhyay2009}). Some of these models involve various general 
relativistic frequencies at preferred radii and the neutron star spin frequency.
Such models can have good predictive power \citep{Bhattacharyya2010}. However,
it is not enough to explain only the frequency in order to understand the kHz QPOs.
The modulation mechanism (how the intensity actually varies) and the
decoherence mechanism (why the QPOs are not very narrow) 
are also required to be understood.

\citet{Mendez2006} proposed that, although the
kHz QPO frequencies are plausibly determined by the characteristic
disk frequencies, the modulation mechanism
is likely associated to the high energy spectral component (e.g.,
accretion disk corona, boundary layer,
etc.). This is because the disk alone cannot explain
the large observed amplitudes, especially at hard X-rays where the contribution
of the disk is small. Moreover, the kHz QPO fractional amplitude increases
with energy, and plausibly reaches a saturation value (e.g., \citet{Gilfanovetal2003}).
Therefore, an accurate measurement of the amplitude vs. energy curve might be
useful to understand the modulation mechanism.

The lack of coherence (aperiodicity) of kHz QPOs could be intrinsic (e.g., shot-noise
models; \citet{vanderKlis2006} and references therein), or might be because of
a decoherence mechanism or frequency drift. The decoherence could be
because of a damped harmonic oscillator or a finite lifetime of a clump of
matter. A timing feature originating from a finite-width disk annulus could
also be broadened because of the superposition of a range of frequencies.
A frequency drift, for example due to a varied preferred disk radius,
might also broaden the feature. Elimination of the effects of frequency superposition
and drift is required to measure the coherence of the underlying signal,
which is essential to understand the kHz QPOs. The shift-and-add technique
of \citet{Mendezetal1998}, which was originally used to discover an upper kHz QPO
from 4U 1608--52, has been used by several authors \citep{Barretetal2005a, Barretetal2005b, 
Barretetal2006, Barretetal2010} to track the frequency drift. Subsequent alignment
of kHz QPOs in smaller time segments make the signal much more prominent,
and allows one to measure its quality (Q) factor or coherence. 
The Q-factor is the ratio of the QPO centroid frequency to the full-width half maximum
(FWHM) of the QPO profile. \citet{Barretetal2005a, 
Barretetal2005b, Barretetal2006} noticed that, while the Q-factor of lower kHz QPOs can be up to 
$\sim 200$, that of upper kHz QPOs is always less than 50.
These authors have also found that the Q-factor of the lower kHz QPO first increases
with frequency, and then after a frequency characteristic of a source, the Q-factor
decreases. They tentatively interpreted this as a
signature of the innermost stable circular orbit (ISCO; but see \citet{Mendez2006}), whose existence is a 
key prediction of strong field general relativity. Note, however, although
the shift-and-add technique tracks the frequency drift correctly, it gives an
average Q-factor value over a long period. Consequently, this technique would
miss a plausible very high Q-factor value, if such a value appears for a short
period of time.

In this Letter, we report the discovery of a very strong lower kHz QPO from the neutron
star LMXB EXO 1745-248, and a very high coherence of this QPO for a short duration. 
%We have also studied the variation of amplitude and coherence with energy of this QPO.

\section{Data Analysis and Results}\label{DataAnalysisandResults}

\begin{figure*}
\centering
\begin{tabular}{c}
\hspace{-1.0cm}
\includegraphics*[width=0.5\textwidth]{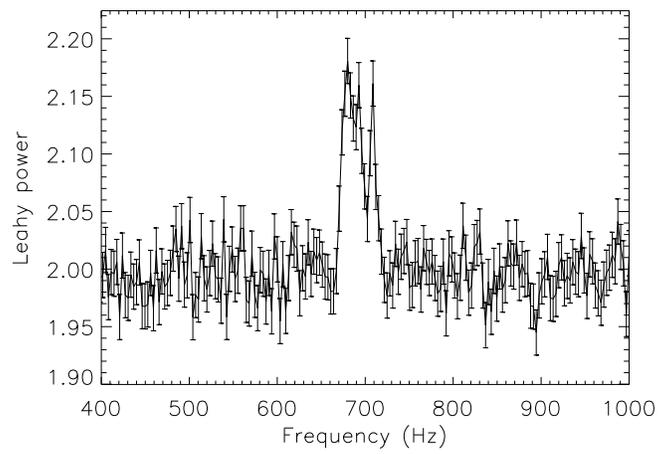}
\end{tabular}
\caption{Power spectrum with frequency resolution of 3.2 Hz from {\it RXTE} PCA data 
(Sep 30, 2000: 15:29:24 to 16:24:09) of the neutron star LMXB EXO 1745--248. 
A kHz QPO at $\sim 690$ Hz with $\approx 20\sigma$ significance is clearly seen 
(\S~\ref{DataAnalysisandResults}).
\label{HF-Powspec1}}
\end{figure*}

The globular cluster Terzan 5 contains several point X-ray sources \citep{Heinkeetal2006}.
Among them the transient neutron star LMXB EXO 1745--248 went into outburst in 2000 and 2002.
These outbursts
were observed with {\it Rossi X-ray Timing Explorer} ({\it RXTE}) between Jul 13
(start time: 13:27:50) and Nov 3 (end time: 00:17:52) in 2000, and
between Jul 2 (start time: 20:38:24) to Jul 22 (end time: 19:42:08) in 2002.
The total observation time was $\approx 144$ ks (proposal nos. P50054, P50138, P70412).
We have searched for kHz QPOs in the corresponding science event mode (resolution 122 $\mu$s)
Proportional Counter Array (PCA) data. In order to do this, we have created one power spectrum
from each segment of continuous observation considering all the PCA channels. Note that
an ObsID may contain more than one segment, and we have removed bursts and data gaps
before doing the Fourier transform. In order to identify a burst, we have calculated average count rate
in each 10 s bin in a segment. We have considered that a bin has a burst, if the count rate of that bin is 
more than 3.2 times that of the previous bin. We have then excluded from 10 s before that bin up to 180 s after that
bin. Using rest of the data,  we have performed Fourier transform on each 10 s bin, and then averaged 
all the Leahy normalized power spectra \citep{vanderKlis1989} within one segment. 
From the entire data set, we have created 146 averaged Leahy power spectra (each from one segment) with resolution
and Nyquist frequency of 0.1 Hz and 2048 Hz respectively. A prominent peak at $691 \pm 25.6$ Hz
has been observed in the Leahy power spectrum of the PCA data of Sep 30, 2000
(15:29:24 to 16:24:09; obsId: 50054-06-11-00; see Fig.~\ref{HF-Powspec1}).
With a frequency resolution of 51.2 Hz, the peak power $\approx 2.103$, implying a single
trial significance of $\approx 1 - 2.68 \times 10^{-95}$ for $326$ time bins of 10 s \citep{vanderKlis1989}. 
A conservative number of trials ($N_{\rm trial}$) can be calculated
from the original number (20480) of powers in a power spectrum multiplied with
the total number (146) of averaged Leahy power spectra searched. With this $N_{\rm trial}$ value,
a kHz QPO can be considered to be detected with a signficance of $ \approx 1 - 8.01 \times 10^{-89} $, i.e., $ \approx 20 \sigma$.
We have then fitted the power spectrum with a
constant+powerlaw+Lorentzian model, taking care of Poissonian noise, red noise and the candidate peak, respectively,
 which gives an acceptable $\chi_{\nu}^2$ (dof) = 0.89 (33).
The Lorentzian describes the kHz QPO, and gives its centroid frequency ($\nu_{\rm QPO}=686.8 \pm 0.9$ Hz), FWHM ($28.3 \pm 2.5$ Hz), 
Q-factor ($24.3 \pm 2.2$) and background corrected fractional rms amplitude ($9.6 \pm 0.3$\%).
 The fractional rms amplitude has been calculated using the integrated power in the Lorentzian component and the standard 
technique mentioned in \citet{vanderKlis1989}. The background subtraction has been done using the equation 1 of \citet{Munoetal2002}.

The very significant kHz QPO is broad and it has sub-features, as mentioned
in our Astronomer's Telegram \citep{MukherjeeBhattacharyya2010}. We have investigated
whether the width of the QPO is a result of decoherence, or caused by
the frequency drift. In order to do this, we have divided the data segment into two
equal parts of 1630 seconds, and created an average power spectrum for each of
them using the same procedure mentioned earlier.
Now the question is whether the frequencies of the kHz QPOs (if detected) of the two 
power spectra are significantly different from each other. First we need to set a detection
criterion. We have considered an $N_{\rm trial} =$ searched frequency range divided by
frequency resolution $= 200/0.1 = 2000$, because (1) the kHz QPO of the entire segment
is already established, and (2) we have searched for the kHz QPO in the half segments
within 100 Hz on each side of $\nu_{\rm QPO}$.
Then a kHz QPO from a half segment is considered to be detected if the product of
$N_{\rm trial}$ and the single trial significance is better than $3\sigma$ with
a frequency resolution of 0.4 Hz or better. The kHz QPOs of both the half segments
have been detected, since the single trial significances are $ \approx 6.70 \times 10^{-32} $
and $ \approx 2.85 \times 10^{-38} $. The best-fit centroid frequency, FWHM, Q-factor and fractional rms amplitude of the
kHz QPOs of first and second half segments are $704.0 \pm 0.9$ Hz, $23.7 \pm 2.5$ Hz, $29.7 \pm 3.2$ and $10.0 \pm 0.3$\%,
and $680.2 \pm 0.3$ Hz, $11.5 \pm 1.0$ Hz, $59.1 \pm 4.9$ and $9.4 \pm 0.3$\% respectively.
Therefore, the Q-factor has increased from that for the entire segment.
Moreover, the significant frequency difference between the two half segments shows that
the structure and the large width of the kHz QPO exhibited in Fig.~\ref{HF-Powspec1} were caused by
frequency drift.

This has motivated us to further divide the data segment in order to check if a higher
Q-factor could be obtained. We have created power spectra using the above mentioned
procedure for quarter segments, one-eighth segments and one-sixteenth segments.
It is difficult to detect the kHz QPO for any one-thirtysecond segment.
The above mentioned criterion allows us to detect kHz QPO in six one-eighth segments
and six one-sixteenth segments. Fig.~\ref{HF-Powspec2} displays that the kHz QPO frequency
shifts from one quarter segment to another. Fig.~\ref{fig_Qfactor-final.ps} shows
that the Q-factor increases for smaller time segments. The highest Q-factor
($264.5 \pm 38.5$) we have obtained is for the sixth one-sixteenth segment
(Fig.~\ref{fig_Qfactor-final.ps}). 
%To the best of our knowledge, this is the largest Q-factor value ever reported for a kHz QPO.
The single trial significance of the kHz QPO of this segment is $ \approx 9.16 \times 10^{-8}$,
which shows that it is somewhat strong 
(panel {\it a} of Fig.~\ref{shiftandaddvs6of16_kHzQPO-final-coma.ps}).

Next, we have checked the effect of the shift-and-add technique (see \S~\ref{Introduction}) on the
narrow kHz QPO features. Panel {\it b} of Fig.~\ref{shiftandaddvs6of16_kHzQPO-final-coma.ps} shows the average power spectrum
from the six one-sixteenth segments with the kHz QPOs aligned at $\nu_{\rm QPO}$.
As expected, the shift-and-add technique makes the kHz QPO more prominent, but
with a Q-factor ($155.5 \pm 10.3$) much smaller than the highest Q-factor ($264.5 \pm 38.5$)
obtained for an individual segment. 

\begin{figure*}
\centering
\begin{tabular}{c}
\hspace{-1.0cm}
\includegraphics*[width=0.5\textwidth]{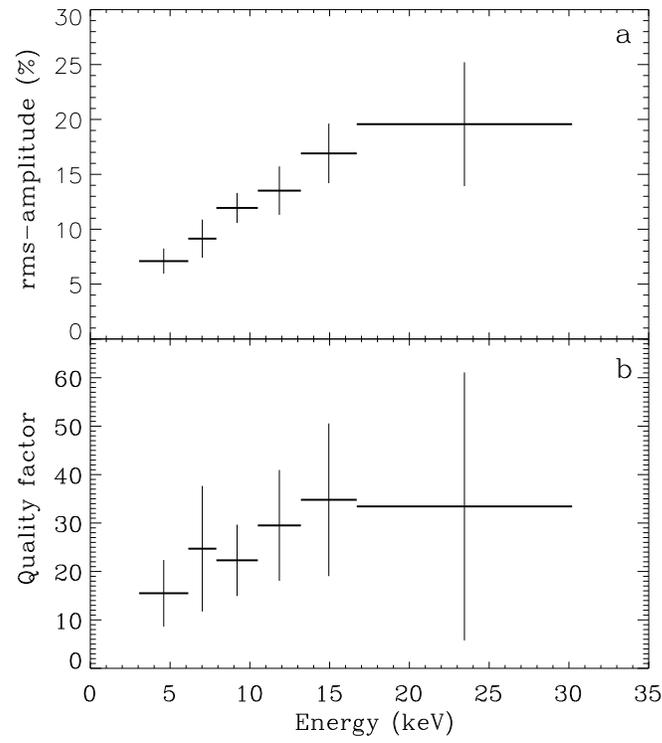}
\end{tabular}
\caption{Properties of the kHz QPO from EXO 1745--248.
Panel {\it a}: shows that the background corrected fractional rms amplitude increases
with energy. 
Panel {\it b}: shows a weak trend of Q-factor increase with energy.
\label{fig_bkgsubs-RMSampandQfactorvsEnergy-final-coma.ps}}
\end{figure*}

Finally, we have plotted the background corrected fractional rms amplitude vs. energy and 
Q-factor vs. energy of the kHz QPO for the entire segment.
(Fig.~\ref{fig_bkgsubs-RMSampandQfactorvsEnergy-final-coma.ps}). The rms clearly
increases with energy. An F-test (null hypothesis probability $\approx 0.01$) between a constant model and a linear model indicates a plausible increase of Q-factor with energy.

\section{Discussion}\label{Discussion}

\begin{figure*}
\centering
\begin{tabular}{c}
\hspace{-1.0cm}
\includegraphics*[width=0.5\textwidth]{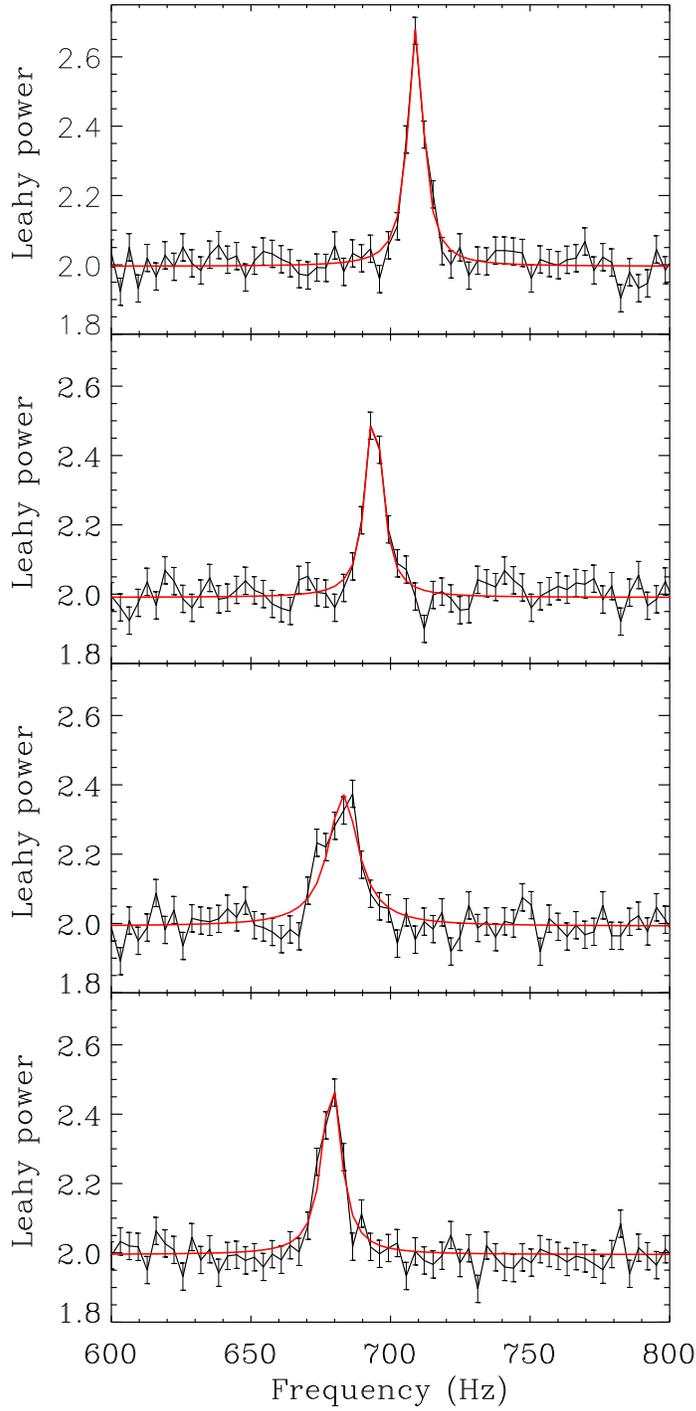}
\end{tabular}
\caption{Power spectra of quarter segments (810 s each; chronologically from
top panel to bottom panel)
of the data set containing the kHz QPO (\S~\ref{DataAnalysisandResults}).
The data points with $1\sigma$ error bars, and the \textquotedblleft 
constant+powerlaw+Lorentzian\textquotedblright model are displayed.
This figure shows that the kHz QPO is very significant in each quarter segment,
and the centroid frequency monotonically decreases with time. 
\label{HF-Powspec2}}
\end{figure*}

\begin{figure*}
\centering
\includegraphics*[width=\textwidth]{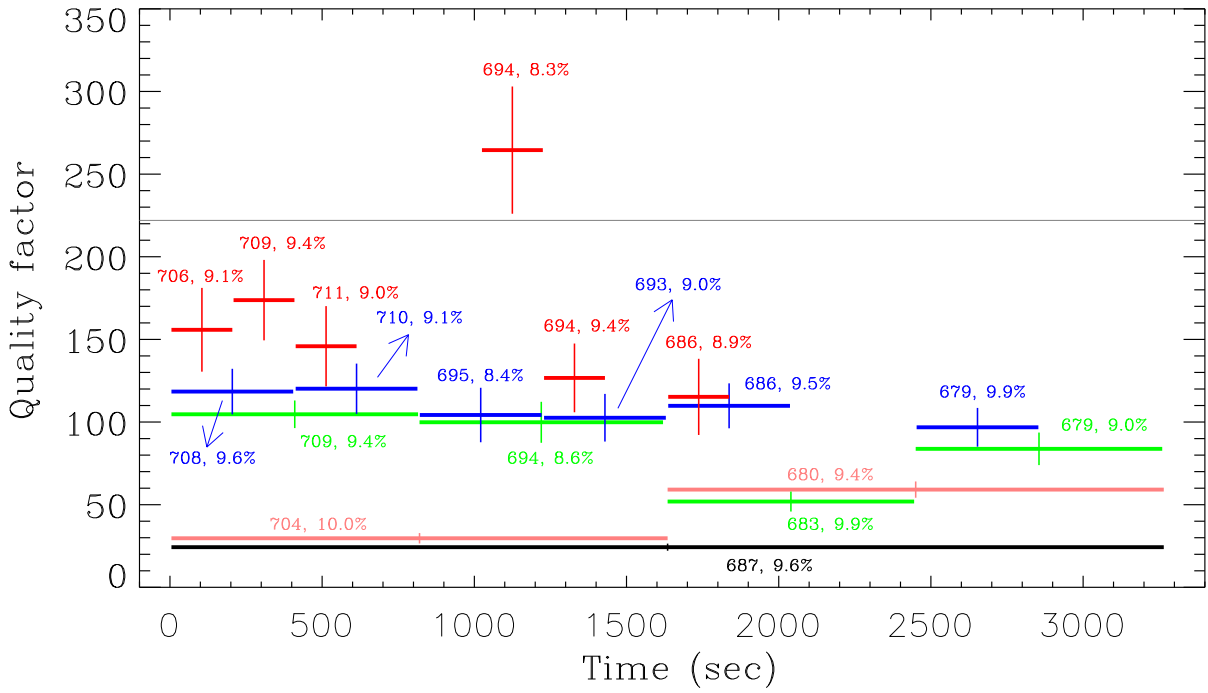}
\caption{Time variation of Q-factor of the kHz QPO from EXO 1745-248 using time segments
of various lengths (\S~\ref{DataAnalysisandResults}). Each horizontal line, with
approximate kHz QPO centroid frequency and rms amplitude written next to it, gives
the time span for which the Q-factor is measured. The corresponding vertical line
gives the $1\sigma$ error of the Q-factor. The light horizontal line gives the 
plausibly highest Q-factor value (222) previously reported (\S~\ref{Discussion}). 
This figure shows a greater Q-factor (264.5), and a significant change in Q-factor at
almost the same frequency.
\label{fig_Qfactor-final.ps}}
\end{figure*}

In this Letter, we report the first detection and analysis of a kHz QPO
from the neutron star LMXB EXO 1745--248. 
%The Q-factor estimated from the entire continuous data segment is less than 50
%(Fig.~\ref{fig_Qfactor-final.ps}), which cannot be used to determine if the feature is
%a lower kHz QPO or an upper kHz QPO (\S~\ref{Introduction}). 
In order to investigate if
its apparent large width and structure are caused by the frequency drift, we have divided the
data segment into smaller parts (see \S~\ref{DataAnalysisandResults}). Note that initially
we have not used the shift-and-add technique, so that the Q-factor of the underlying
signal does not get averaged, and we can study its time variation within a short 
duration. The measured $Q > 50$ values for all the smaller parts except one
(Fig.~\ref{fig_Qfactor-final.ps}) show that this is a lower kHz QPO 
(\S~\ref{Introduction}). We have not found an upper kHz QPO even by the 
shift-and-add technique (\S~\ref{DataAnalysisandResults}). The frequency 
of the QPO roughly monotonically drifts by $\sim 30$ Hz in about an hour
(Fig.~\ref{HF-Powspec2}). The variations of fractional rms amplitude ($8-10 \%$) and Q-factor, 
however, are not correlated with the frequency change in our limited time and frequency
ranges.

The fact that the Q-factor is generally larger for smaller time segments, and the frequency
shifts from one segment to another shows that the frequency drift plays a major role
in broadening the QPO. At the level of one-sixteenth segment (200 s;
\S~\ref{DataAnalysisandResults}), the width of the
QPO may still be caused by the combined effect of a small frequency drift, the superposition of
a small frequency range, and a decoherence mechanism. Since we cannot divide the time
segment any further (see \S~\ref{DataAnalysisandResults}), our estimated Q-factor
is only a lower limit. The corresponding lower limit of the coherence time $\tau$ can be 
calculated from $\tau = 1/(\pi\times{\rm FWHM})$, 
where the QPO profile is fitted with a Lorentzian,
and we assume that the signal consists of exponentially damped sinusoidal oscillator \citep{Barretetal2005a, vanderKlis2006}. 

\begin{figure*}
\centering
\includegraphics*[width=0.5\textwidth]{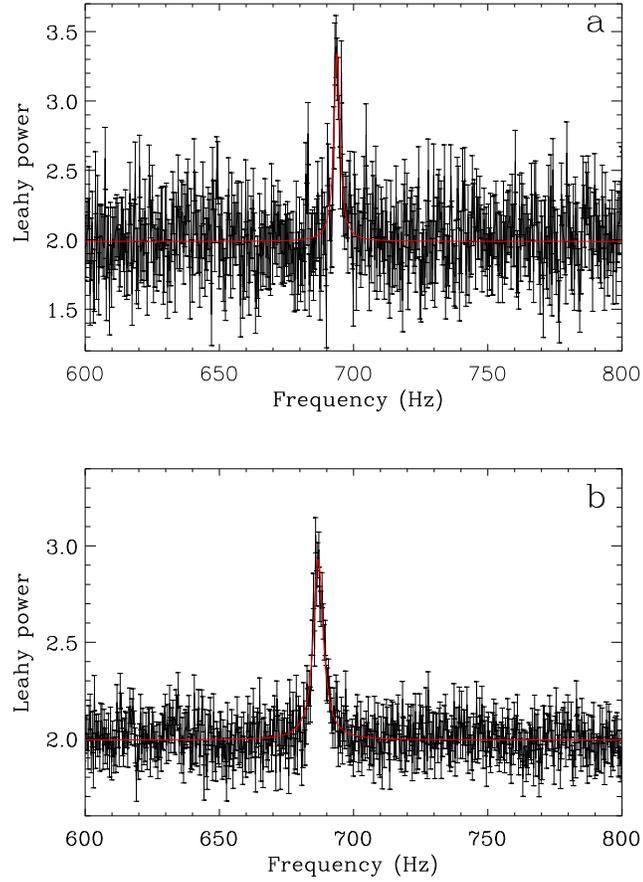}
\caption{Power spectra of the data set from EXO 1745--248 containing the kHz QPO. 
Each panel shows the data points with $1\sigma$ error bars, and the \textquotedblleft
constant+powerlaw+Lorentzian\textquotedblright model.
Panel {\it a}: shows the kHz QPO of the one-sixteenth segment with a Q-factor of 
$264.5 \pm 38.5$ (see Fig~\ref{fig_Qfactor-final.ps} and \S~\ref{DataAnalysisandResults}).
Panel {\it b}: shows the kHz QPO after applying the shift-and-add technique to six
one-sixteenth segments (\S~\ref{DataAnalysisandResults}).
This figure shows that while the shift-and-add technique makes the QPO more prominent,
the measured coherence from this technique can be much lower than the highest coherence
obtained from an individual segment.
\label{shiftandaddvs6of16_kHzQPO-final-coma.ps}}
\end{figure*}

For a one-sixteenth segment we have measured a Q-factor of $264.5 \pm 38.5$,
which might be the largest value reported in the literature (\citet{Barretetal2005b} reported
$Q = 222\pm24$ for 4U 1636--536). 
Moreover, this value appears at $\approx 694$ Hz. 
From \citet{Barretetal2006} we find that Q-factor near this frequency is found to be less than 150.
The minimum $\tau$ corresponding to a Q-factor of 264.5 is 0.12 s. 
Many models will find it difficult to explain such a large $\tau$ (see \citet{Barretetal2005a}
for a discussion). For example, kHz QPO models based on accretion flow  
inhomogeneities in the form of clumps
should have $\tau \lsim 0.01$ s \citep{Barretetal2005a}. Therefore, an effective 
way to constrain and rule out kHz QPO models is to observationally 
find high Q-factors, even for a short
duration. If the Q-factor considerably varies within short ranges of time and frequency (see
Fig.~\ref{fig_Qfactor-final.ps}), the shift-and-add technique (\S~\ref{Introduction}), 
which gives an average Q-factor and QPO structure, cannot find such high Q-factors.
We have demonstrated this in Fig.~\ref{shiftandaddvs6of16_kHzQPO-final-coma.ps} 
(\S~\ref{DataAnalysisandResults}).
Therefore, while the Q-factor vs. frequency trend reported by Barret and coauthors 
should be useful to probe fundamental physics, the Q-factor of lower kHz QPOs can be much
larger at times than the reported peak values of these authors.

Finally, the observed increase of fractional rms amplitude with energy is common for kHz QPOs
(e.g., \citet{Gilfanovetal2003}), and may provide crucial information about the hard X-ray
modulation mechanism by identifying a modulation site (see \S~\ref{Introduction}).
Moreover, a plausible increase of Q-factor with energy indicates that oscillations from 
harder X-ray components are more coherent than those from softer X-ray components.
{\it Astrosat} with its sufficient time resolution and an area larger than {\it RXTE} PCA
at hard X-rays might be ideal to probe these trends.

%\section{Summary}\label{Summary}

%(1) We report the first detection of a kHz QPO from the neutron star LMXB EXO 1745--248.
%	This is a lower kHz QPO with a significance of ${\bf number}\sigma$.
 
%(2) For a duration of 200 s, the Q-factor of the kHz QPO is $264\pm${\bf number} corresponding
%	to a coherence time of ${\bf number}$ s. To the best of our knowledge, this is
%	the most coherent kHz QPO reported, and should have implications for the theoretical
%	models.

%(3) We conclude that the coherence of a kHz QPO can vary significantly in a timescale of
%	$\lsim 200$ s. Therefore, there can be durations in which the coherence is much greater
%	than the value estimated using the shift-and-add technique.

%\section*{Acknowledgments}

\acknowledgments

We thank Deepto Chakrabarty for a discussion, and an anonymous referee for constructive comments
 which improved the Letter. SB thanks ISSI for their hospitality. This work was supported in part
 by the TIFR plan project 11P-408 (P.I.:K. P. Singh).

{}

\clearpage

\clearpage

\clearpage

\clearpage

\clearpage

\end{document}